         \documentclass[twocolumn,eqsecnum,aps,prb,floatfix]{revtex4}
         \usepackage[dvips]{graphicx}
\providecommand{\mB}{\mu_{\rm B}}
\providecommand{\veck}{\mathbf{k}}
\providecommand{\ve}{\varepsilon}
\providecommand{\vecQ}{\mathbf{Q}}
\providecommand{\vecG}{\mathbf{G}}
\providecommand{\vecq}{\mathbf{q}}
\providecommand{\vecr}{\mathbf{r}}
\providecommand{\murakami}
{Murakami98a,Murakami98b,murakami99a,murakami99b,murakami99c,murakami00b}

\begin{document}
\title{Magnetic Resonant X-Ray Scattering in KCuF$_3$}
\author{Manabu Takahashi}
\affiliation{Faculty of Engineering, Gunma University, Kiryu, Gunma 376-8515, Japan}
\author{Manabu Usuda}
\author{Jun-ichi Igarashi}
\affiliation{Synchrotron Radiation Research Center, Japan Atomic Energy Research Institute,\\ Mikazuki, Sayo, Hyogo 679-5148, Japan}
\date{\today}
%
%--------%---------%---------%---------%---------%---------%---------%---------%
%
\begin{abstract}
We study the magnetic resonant x-ray scattering (RXS) spectra around the $K$
 edge of Cu in KCuF$_3$ on the basis of an {\em ab initio} calculation.  We use
 the full-potential linearlized augmented plane wave method in the LDA$+U$
 scheme, and introduce the lattice distortion as inputs of the calculation.
  We obtain finite intensity on magnetic superlattice spots, about three
 orders of magnitude smaller than on orbital superlattice spots, by taking
 account of the spin-orbit interaction (SOI).  No intensity appears without
 the SOI, indicating that the intensity arises not from the spin
 polarization but from the orbital polarization in $4p$ states.  The present
 calculation reproduces well the experimental spectra as functions of photon
 energy and of azimuthal angle.  We also calculate the RXS intensity on
 orbital superlattice spots.  It is found that the intensity increases with
 increasing Jahn-Teller distortion. The spectra remain nearly the same in
 the nonmagnetic state given by the simple LDA, in which the orbital
 polarization in the $3d$ states is much smaller.  This strongly suggests
 that the intensity on orbital spots is mainly controlled by the lattice
 distortion, not by the $3d$ orbital order itself. 
\end{abstract} 


\maketitle

% 
%--------%---------%---------%---------%---------%---------%---------%---------% 
%\narrowtext

\section{introduction}

The resonant x-ray scattering (RXS) has attracted much attention, since the
 orbital order can be directly probed by the RXS signals on the orbital
 superlattice spots.  Several RXS experiments have been carried out on
 typical perovskite compounds, \cite{\murakami,Nakao00} in which an
 orbitally ordered state may be stabilized below a critical
 temperature.\cite{Kanamori,Khomskii72} For the $K$ edge in transition-metal
 compounds, RXS intensities arise from the modulation of $4p$ states in the
 intermediate states of the resonant process.  Since $4p$ states are not the
 states of orbital ordering, this causes complications on the interpretation
 of RXS intensities.

A mechanism was proposed that the modulation of $4p$ states comes from
 orbitally polarized $3d$ states on the Mn$^{3+}$ ion through the
 anisotropic terms of the $3d$--$4p$ Coulomb interaction in
 LaMnO$_3$,\cite{Ishihara,Ishihara2} and thereby it was argued that the RXS
 intensities on the superlattice spots are a direct reflection of orbital
 order. On the other hand, subsequent studies based on band structure
 calculations\cite{Elfimov,Benfatto,Taka1} have revealed that $4p$ states of
 Mn$^{3+}$ ion so extend in space that they are considerably modified
 by neighboring electronic states through the Jahn-Teller distortion (JTD).
  It has been concluded that the effect of the JTD on the RXS spectra is
 much larger than that of the $3d$-$4p$ Coulomb interaction.

RXS experiments were carried out also on YVO$_3$[\onlinecite{murakami00b}] and
 YTiO$_3$[\onlinecite{Nakao00}]. Since the JTD is considerably smaller in these
 materials than in LaMnO$_3$, one may think the effect of the Coulomb
 interaction important.  This is not the case, though; {\em ab initio}
 calculations indicated that the effect of lattice distortion on RXS
 intensities is much larger than that of the $3d$-$4p$ Coulomb interaction
 even in these materials.  \cite{Taka2,Taka3} The calculation was carried out
 within the muffin-tin (MT) approximation on the lattice parameters
 determined from the experiment.  Since the MT approximation averages the
 potential coming from orbitally polarized $3d$ states, it works to
 eliminate the effect of the anisotropic terms of the $3d$-$4p$ Coulomb
 interaction on the $4p$ states.  We found that the spectra consist of
 several peaks, to which the GdFeO$_3$-type distortion and the JTD
 contribute differently, in good agreement with the experiment.

Recently, an RXS experiment  has been carried out on KCuF$_3$, in which the
 intensities on the magnetic and orbital superlattice spots have been
 reported.\cite{Paolasini,Caciuffo} The magnetic RXS has been studied for
 typical materials such as CoO[\onlinecite{Neubeck1}] and NiO[\onlinecite{Neubeck2}].
  However, the analyses of the experimental data are limited to model
 calculations.\cite{Iga1,Iga2} As regards the intensities on the orbital
 spots, they increase with going down through the Ne\'el temperature in the
 experiment, and are interpreted as a consequence of the increase of the $3d$
 orbital order parameter in the magnetic phase.\cite{Paolasini} This
 interpretation seems obscure unless it is clarified how the $3d$ orbital
 order controls the RXS spectra.  The purpose of this paper is to
 elucidate the mechanism of the RXS spectra on the orbital and magnetic
 superlattice spots through {\em ab initio} calculations.  We use the
 full-potential linearlized augmented plane wave (FLAPW) method in the
 LDA$+U$ scheme, where the local $3d$--$3d$ Coulomb interaction is introduced on Cu
 sites.\cite{Anisimov}

We obtain the RXS intensity on magnetic spots by taking account of the
 spin-orbit interaction (SOI).  The calculation has to be accurate, since
 the intensities are about three orders of magnitudes smaller than those on
 orbital spots.  This seems the first case to have evaluated the magnetic
 RXS spectra in the {\em ab initio} level.  We reproduce well the
 experimental spectra as functions of photon energy and of azimuthal angle.
  No intensity appears without the SOI, indicating that the intensity arises
 from the orbital polarization, not from the spin polarization, in the $4p$
 states.  On the other hand, the spectra change little with turning off the
 SOI from the $3d$ states. Since the $3d$ orbital moment disappears in this
 condition, the $3d$ orbital moment plays no role to induce the orbital
 polarization in $4p$ states.  This shows a contrast with the magnetic RXS
 in NiO and CoO, where the $3d$ orbital moment plays an important role
 through the mixing of the $4p$ states with the $3d$ states of neighboring
 Cu sites and the intra-atomic $4p$-$3d$ Coulomb interaction.

We also calculate the RXS spectra on the orbital superlattice spots.  A
 similar calculation has already been carried out.\cite{Caciuffo} We come to
 the same conclusion that the RXS intensity is mainly controlled by lattice
 distortion.  We add several aspects of making clear the mechanism by
 calculating the spectra with changing magnitude of the JTD and in various
 magnetic phases.  We explicitly show that the magnetic order itself has
 little influence on the intensity and that the intensity increases with
 increasing JTD strength.  In this respect, the increase of the intensity
 with passing through the Ne\'el temperature has to accompany a considerable
 increase of the JTD strength in the magnetic phase.  We also predict a new
 azimuthal angle dependence different from the previously assumed one but
 consistent with the experiment.\cite{Caciuffo}

In Sec.~II, we study electronic structures in various phases with the FLAPW
 method in the LDA$+U$ scheme.  In Sec~III, the absorption spectra and the
 RXS spectra are presented in comparison with the experiment.  Section IV is
 devoted to concluding remarks.

\section{electronic structures}
%--------+---------+---------+---------+---------+---------+---------+---------+
\begin{figure}
\includegraphics[width=7cm,keepaspectratio]{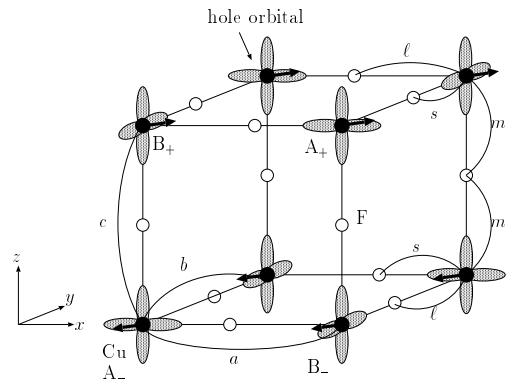}
\caption{Sketch of the ordering pattern of KCuF$_3$.
Orbitals $\psi_{x^2-z^2}$ and $\psi_{y^2-z^2}$ are
antiferro-orbitally ordered with the ordering vector
$\vecQ_{\rm orb}=(\pi/a,\pi/b,\pi/c)$. Arrows represent the direction of
magnetic moments with the ordering vector $\vecQ_{\rm mag}=(0,0,\pi/c)$.
Small solid spheres represent F atoms.
K atoms are omitted in the figure.
}
\label{fig:crys}
\end{figure}
%--------+---------+---------+---------+---------+---------+---------+---------+
KCuF$_3$ has a typical perovskite structure, and undergoes a structural
 phase transition at $T\approx 800$ K, below which the JTD and the
 associated antiferro-orbital (AFO) long-range order emerges with the
 ordering vector $\vecQ_{\rm orb}=(\pi/a,\pi/b,\pi/c)$. Figure \ref{fig:crys} shows the
 schematic view of the ordering pattern.  It is tetragonal with $a=b=4.141$\AA,
 $c=3.924$\AA\ at room temperature; three different bond lengths are
 given by $\ell=2.253$\AA, $s=1.888$\AA, $m=1.962$\AA.\cite{Buttner,Hutings}
 Since the lattice parameters are not known at
 low temperatures, we use these values in the following calculation.  There
 are two inequivalent Cu sites in a unit cell, which will be called as A and
 B sites.  With further decreasing temperatures, an antiferromagnetic (AFM)
 long-range order develops below 38 K with the ordering vector $\vecQ_{\rm
 mag}=(0,0,\pi/c)$.  Therefore four inequivalent Cu sites, A$_+$, A$_-$,
 B$_+$, B$_-$, exist.  The local magnetic moment is pointing to either the
 $(110)$ or $(1\bar1 0)$ direction with the value 0.48 $\mB$.  The recent
 magnetic scattering experiment gives the ratio of the orbital angular
 momentum ($L$) to the spin angular momentum ($S$) as
 $L/S=0.29$.\cite{Caciuffo}

We carry out the band structure calculation in the LDA$+U$ scheme by setting
 the parameters of the intraatomic Coulomb interaction between $3d$ orbitals
 such that $U=3.0$ eV and $J=1.0$ eV.  We take account of the SOI to deal
 with the orbital moment, which is important for the RXS spectra on the
 magnetic spots.  First assuming the AFM long-range order, we obtain an
 orbital long-range order with the ordering vector $\vecQ_{\rm orb}$; the
 {\em hole} numbers inside the Cu MT sphere in the $3d$ states $\varphi_{x^2-z^2}$
 ($\varphi_{y^2-z^2}$) and $\varphi_{3y^2-r^2}$ ($\varphi_{3x^2-r^2}$) are
 given by $n_{x^2-z^2}(n_{y^2-z^2})=0.65$ $n_{3y^2-r^2}(n_{3x^2-r^2})=0.08$
 in A$_\pm$ (B$_\pm$) sites, respectively.  Note that the difference in the
 values between A$_+$ and A$_-$ sites and B$_+$ and B$_-$ sites is
 negligibly small. The local magnetic moment is obtained as $0.79\mB$
 with $S=0.34$, $L=0.11$, $L/S=0.33$, which is somewhat larger than the
 experimental one.  In order to see the effect of the magnetic order on
 orbital polarization, we carry out the same LDA$+U$ calculation in the
 presence of the ferromagnetic (FM) long-range order. We obtain the same
 type of orbital order, in which $n_{x^2-z^2}(n_{y^2-z^2})=0.66$
 $n_{3y^2-r^2}(n_{3x^2-r^2})=0.10$ in A (B) sites, respectively.  Since
 the hole numbers are similar to that of the antiferromagnetic state, we
 conclude that the magnetic order has little influence on the orbital order
 parameter. On the other hand, the orbital polarization is rather sensitive
 to the intraatomic Coulomb interaction.  The simple LDA ($U=0$) gives the
 orbital polarization much smaller than the values mentioned above, that is,
 $n_{x^2-z^2}(n_{y^2-z^2})=0.26$ and $n_{3y^2-r^2}(n_{3x^2-r^2})=0.11$ per
 spin in the A(B) sites in the nonmagnetic (NM) phase.

%--------+---------+---------+---------+---------+---------+---------+---------+
\begin{figure}
\includegraphics[width=7cm,keepaspectratio]{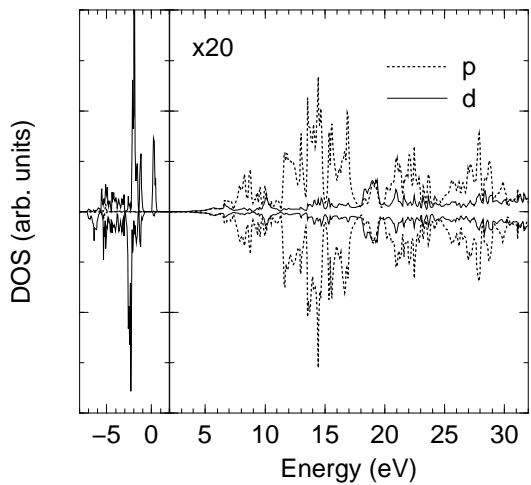}
\caption{Density of states projected on the $s$, $p$ and $d$
symmetric states inside the Cu muffin-tin sphere,
calculated in the AFM phase of the LDA$+U$ scheme.
The origin of energy is the bottom of the conduction band.}
\label{fig:doscalc}
\end{figure}
%--------+---------+---------+---------+---------+---------+---------+---------+
Figure \ref{fig:doscalc} shows the calculated density of states (DOS)
 projected on the $d$ and $p$ symmetries of a Cu site, which is defined
 inside the MT sphere, in the AF phase given by the LDA$+U$ method.  The
 origin of the energy is the bottom of the conduction band.  The DOS projected
 onto the $d$ symmetric states is almost concentrated in the region of
 energy less than 2 eV. It has a finite gap $\sim 0.82$ eV. The left part of
 DOS is shown in a 20 times magnified scale.  The most part of probability
 distributes in the interstitial region or on neighboring MT spheres, and
 thereby the magnitude of the DOS projected onto the $p$ symmetry becomes
 much smaller than the $3d$ DOS.  We refer to the $p$ symmetric states as
 $4p$ states, because the wave functions are quite close to the atomic $4p$
 wave functions within the MT sphere.  Of course the ``4p" states lose the
 atomic character outside the MT sphere, forming an energy band with its
 width being as large as $30$ eV.  The $4p$ DOS shows almost no exchange
 splitting in spite of the magnetic ground state, indicating that the $3d$
 states have little influence on $4p$ states. 

\section{Absorption Coefficient and RXS spectra}

We define the $1s$--$4p$ dipole transition density matrix 
$\tau^{(j)}_{mm'}(\ve)$ assigned to each Cu site as
\begin{eqnarray}
\tau^{(j)}_{mm'}(\ve) &=& \sum_{n,\veck}\int r^2 dr\, r'^2 dr' [R_{j1s}^*(r ) r  {\cal P}^{(j)}_m\phi_{n,\veck}(r )]^* \nonumber \\
                &\times& [R_{j1s}^*(r') r' {\cal P}^{(j)}_{m'}\phi_{n,\veck}(r')] \delta(\ve-\ve_{n,\veck}),
\label{transition}
\end{eqnarray}
where $j$ stands for sublattices A$_\pm$ or B$_\pm$. Suffixes $m$, $m'$
represent the $x$, $y$, and $z$ axes, 
which are chosen to be parallel to the $a$, $b$, and $c$ axes,
respectively (see Fig. \ref{fig:crys}). 
The $\phi_{n,\veck}$ represents the wave function with the band index
$n$, wave-vector $\veck$ and energy $\ve_{n,\veck}$ which is
larger than the Fermi energy. Operator ${\cal P}^{(j)}_m$ projects the wave
function $\phi_{n,\veck}$ on the $m$ ($m\in x,y,z$) component of the
$p$ symmetric part on the Cu $j$ site. The $R_{j1s}$ represents the Cu
$1s$ wave function on the Cu $j$ site. We evaluate this density matrix using the FLAPW
wave function.

\subsection{Absorption Coefficient}

\begin{figure}
\includegraphics[width=7cm,keepaspectratio]{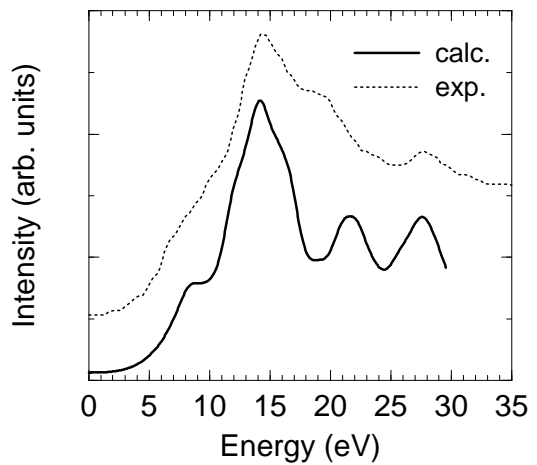}
\caption{
Absorption coefficient $A(\omega)$ as a function of photon energy
around the $K$ edge of Cu, in comparison with the experimental data of 
fluorescence (Ref. \onlinecite{Caciuffo}).
The core-hole energy is adjusted such that the calculated peak position 
coincides with the experimental one.
}
\label{fig:absorp}
\end{figure}

We first discuss the absorption
coefficient $A(\omega)$ around the $K$ edge.
Neglecting the core-hole potential working on the $4p$ states
in the final state of the absorption process, 
we have the expression for $A(\omega)$,
\begin{eqnarray}
&&A(\omega)\propto \sum_{n,\veck}\left|\int r^2 dr R_{1s}^*(r ) r  
 {\cal P}_m\phi_{n,\veck}(r )\right|^2 \delta(\omega-\ve_{n,\veck}-\ve_{1s})
 \nonumber \\
 &&= \tau^{(j)}_{xx}(\omega-\ve_{1s})+\tau^{(j)}_{yy}(\omega-\ve_{1s})
                  +\tau^{(j)}_{zz}(\omega-\ve_{1s}) ,
\label{eq.abs}
\end{eqnarray}
where $\ve_{1s}$ is the energy of the $1s$ state.
The dependence on sublattices can be neglected.
%Since the factor before the $\delta$-function in Eq.~(\ref{eq.abs})
%is nearly independent of $n$ and $\veck$,
$A(\omega)$ is nearly proportional to the $4p$ DOS.
Figure \ref{fig:absorp} shows the calculated result
in comparison with the experiment,
where $\ve_{1s}$ is adjusted such that 
the position of the main peak coincides with the experimental one.
It is difficult to determine the absolute value of $\ve_{1s}$ 
within the present status of the {\em ab initio} calculation, 
since several screening processes are neglected.
The spectral shape is found in good agreement with the experimental curve.
If the core-hole potential would be taken into account,
the intensities in low frequency region may be further enhanced.
%--------+---------+---------+---------+---------+---------+---------+---------+
%--------+---------+---------+---------+---------+---------+---------+---------+

\subsection{RXS Spectra}
%--------+---------+---------+---------+---------+---------+---------+---------+
\begin{figure}
\includegraphics[width=7cm,keepaspectratio]{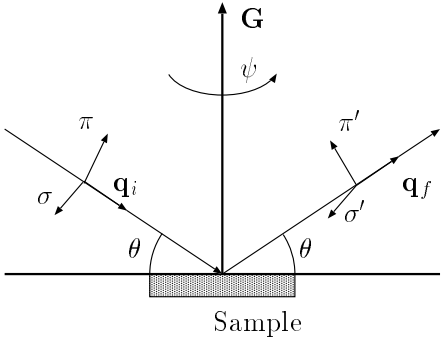}
\caption{RXS geometry. Incident photon with wave vector $\vecq_i$ 
and polarization $\sigma$ or $\pi$ is scattered into the state with
wave vector $\vecq_f$ and polarization $\sigma'$ or $\pi'$ at Bragg angle
$\theta$. The sample crystal is rotated by azimuthal angle $\psi$ around
the scattering vector $\vecG$.
}
\label{fig:geometry}
\end{figure}
%--------+---------+---------+---------+---------+---------+---------+---------+

The conventional RXS geometry is shown in Fig.~\ref{fig:geometry};
photon with frequency $\omega$, momentum $\vecq_i$,
polarization $\mu$ ($=\sigma$ or $\pi$) is scattered into state with
momentum $\vecq_f$ and polarization $\mu'$ ($=\sigma'$ or $\pi'$).
The scattering vector $\vecG$ is defined by $\vecq_f-\vecq_i$.
In such a situation, we have the RXS intensity $I(\vecG,\omega)$ 
\begin{equation}
 I(\vecG,\omega)\propto \left| \sum_{mm'}E_m^{\rm out}
        M_{mm'}(\vecG,\omega)E_{m'}^{\rm in}\right|^2,
\end{equation}
with 
\begin{eqnarray}
&& M_{mm'}(\vecG,\omega) = \nonumber \\
&&\frac{1}{\sqrt{N}}\sum_{j,\Lambda}
  \frac{\langle g|x_m(j)|\Lambda\rangle
        \langle\Lambda|x_m'(j)|g\rangle}
       {\hbar\omega-(E_{\Lambda}-E_g)+i\Gamma}\exp(-i\vecG\cdot\vecr_j). \nonumber \\
\end{eqnarray}
Here $E^{\mbox{\scriptsize in(out)}}_x$ is the $x$ component 
of polarization vector of the incident (scattered) photon,
and $j$ runs over Cu lattice sites. 
The $|g\rangle$ represents the ground state with energy $E_g$. 
The $|\Lambda\rangle$ represents the intermediate state with energy $E_\Lambda$;
it consists of the excited electron on the $4p$ states and a hole 
on the $1s$ state. $\Gamma$ describes the broadening due to the $1s$ 
core-hole lifetime. The RXS spectra are rather insensitive with 
varying $\Gamma$ values; we set ${\Gamma}=1$ eV.
The dipole operators $x_\alpha(j)$ at site $j$ are defined as
$x_1(j)=x$, $x_2(j)=y$, $x_3(j)=z$ in the coordinate frame fixed to
the crystal axes with the origin located at the center of site $j$.
Hereafter $\vecG$ is expressed in unit of $(2\pi/2a,2\pi/2b,2\pi/2c)$.
Just like the absorption coefficient, we neglect the core-hole potential 
in the intermediate state. Then the amplitude $M_{mm'}(\vecG,\omega)$
is expressed in terms of the density matrices:
\begin{eqnarray}
&& M_{mm'}(\vecG,\omega) = \nonumber \\
&& \frac{1}{\sqrt{N}}\sum_j \exp(-i\vecG\cdot\vecr_j)
   \int \frac{\tau^{(j)}_{mm'}(\ve){\rm d}\ve}
             {\hbar\omega-\ve+\ve_{1s}+i\Gamma}.
\label{eq.amp}
\end{eqnarray}

\subsubsection{on Magnetic Superlattice Spots}
%--------+---------+---------+---------+---------+---------+---------+---------+
\begin{figure}
\includegraphics[width=7cm,keepaspectratio]{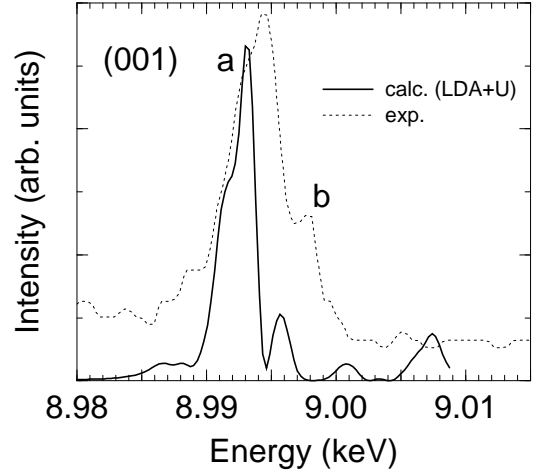}
\caption{Calculated RXS spectra on a magnetic superlattice spot $(001)$
in comparison with the experiment (Ref. \onlinecite{Caciuffo}).
The core-hole energy is set the same as in the case of absorption coefficient
(Fig. \ref{fig:absorp}).
}
\label{fig:RXSmag}
\end{figure}
%--------+---------+---------+---------+---------+---------+---------+---------+
For the magnetic superlattice spots $\vecG=(00m)$
with $m$ being odd integers, Eq.(\ref{eq.amp}) contains the following
combination of density matrices:
\begin{eqnarray}
&& \tau^{({\rm A}_+)}(\ve)- \tau^{({\rm A}_-)}(\ve)
      + \tau^{({\rm B}_+)}(\ve)-\tau^{({\rm B}_-)}(\ve) \nonumber \\
&&=\left( \begin{array}{rrr} 0           & 0 &\eta(\ve)  \\
                                       0           & 0 &\eta(\ve) \\
                                       -\eta(\ve)  &-\eta(\ve)    & 0
                    \end{array} \right).
\end{eqnarray}
Only the off-diagonal elements  survives with an antisymmetric form.
\cite{Iga1,Iga2}
They would vanish without the SOI, indicating that they arise from
the {\em orbital polarization}\footnote{
The orbital polarization in the $4p$ states here means the polarization
related to imaginary wavefunctions, which gives rise to orbital moment.
On the other hand, the $4p$ orbital polarization for the case of orbital 
superlattice spots means the polarization with respect to real wavefunctions 
such as $x$, $y$, $z$ types.}
in the $4p $ states.
From this expression, Eq.(\ref{eq.amp}) becomes 
\begin{equation}
 I(\vecG,\omega) \propto \left| 
        (C_{yz}+C_{zx})\int d\ve\frac{\eta(\ve)}
          {\omega-\ve+\ve_{1s}+{\rm i}{\Gamma}}\right|^2,
\label{eq.RXSmag}
\end{equation}
with $C_{ij}= E^{\rm out}_i E^{\rm in}_j  - E^{\rm out}_j E^{\rm in}_i$.
Since the polarization dependent part $C_{yz}+C_{zx}$ is factored out,
the photon energy dependence is independent of polarization.

The actual calculation in the LDA$+U$ scheme is rather heavy, since the
 number of atoms in the unit cell becomes as large as 20 in the AFM phase.
  Highly accurate calculations are required, since the intensity is about
 three orders of magnitude smaller than that for the orbital superlattice
 spots, as shown below.  It is not easy to sum up many k-points in
 evaluating $\eta(\ve)$, though.  Fortunately, summing $8\times 8\times 8$
 k-points in the first Brillouin zone seems sufficient in comparison with
 the result for summing $6\times 6\times 6$ k-points.  Figure
 \ref{fig:RXSmag} shows the calculated spectra for $\vecG=(001)$ as a
 function of photon energy in comparison with the experiment.
  \cite{Caciuffo}  The intensity in the the $\sigma\to\sigma'$ channel does
 not appear in agreement with the experiment.  Its spectral shape consists
 of several peaks.  The highest and next highest peaks correspond well to
 peaks a and b, although their energies are slightly shifted.  Note that the
 core-hole energy has already been adjusted such that the calculated
 absorption peak coincides with the experimental one.  We have checked that
 the spectra change little with turning off the SOI from the $3d$ states.
 This suggests that the $3d$ orbital moment has little influence on inducing
 the $4p$ orbital polarization. 

\begin{figure}
\includegraphics[width=7cm,keepaspectratio]{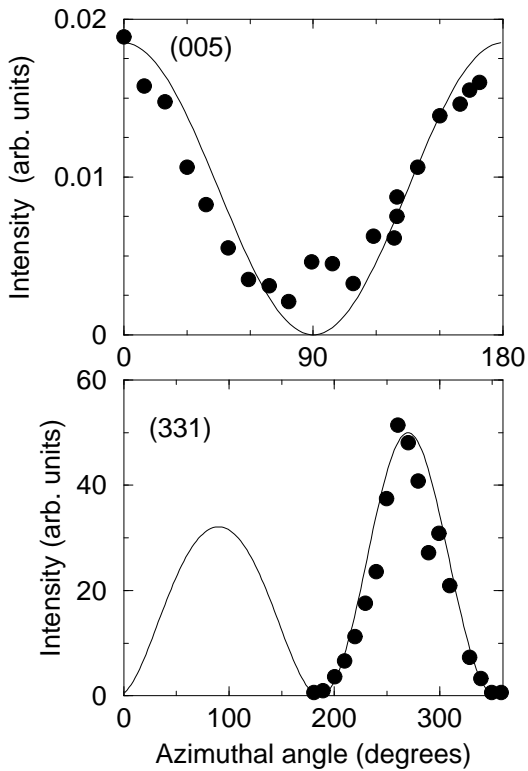}
\caption{Azimuthal angle dependence of the peak intensity
on a magnetic superlattice spot $(005)$ and on an orbital superlattice 
spot $(331)$, in comparison with the experiment (Ref. \onlinecite{Caciuffo}).
} 
\label{fig:azim} 
\end{figure}

From Eq.~(\ref{eq.RXSmag}), we derive explicitly 
the azimuthal angle dependence,
\begin{eqnarray}
 I(\vecG,\omega) &\propto& 0 \quad 
 {\rm for}\ \sigma\to\sigma'    \nonumber \\
                   &\propto& |\cos\theta\cos\psi|^2 \quad
 {\rm for}\ \sigma\to\pi', 
\end{eqnarray}
where $\sin\theta=0.088$ for the $(001)$ spot and $\sin\theta=0.439$ 
for the $(005)$ spot.
Here the azimuthal angle $\psi=0$ is defined such that the $(\bar110)$ axis is contained
in the scattering plane. Figure \ref{fig:azim} shows the peak intensity
on $(005)$ as a function of $\psi$, in comparison with the experiment.
The dependence agrees with the experiment.
%--------+---------+---------+---------+---------+---------+---------+---------+
%--------+---------+---------+---------+---------+---------+---------+---------+

\subsubsection{on Orbital Superlattice Spots}

For the orbital superlattice spots $\vecG=(h\ell m)$ 
with $h$, $l$, and $m$ being odd integers, Eq.~(\ref{eq.amp}) contains
the following combination of density matrices:
\begin{eqnarray}
&& \tau^{({\rm A}_+)}(\ve)+ \tau^{({\rm A}_-)}(\ve)
     -\tau^{({\rm B}_+)}(\ve)-\tau^{({\rm B}_-)}(\ve) \nonumber \\
&&\approx\left( \begin{array}{rrr} \alpha(\ve)-\beta(\ve)  &0  &\gamma(\ve) \\
                             0   &\beta(\ve)-\alpha(\ve)  &\delta(\ve)       \\
                        -\gamma(\ve) & -\delta(\ve) & 0
                    \end{array} \right).
\end{eqnarray}
The off-diagonal elements are about two orders of magnitude smaller than
the diagonal terms.
Neglecting such terms, we obtain
\begin{equation}
I(\vecG,\omega) \propto \left|C_{x^2-y^2} 
        \int d\ve\frac{\alpha(\ve)-\beta(\ve)}
          {\omega-\ve+\ve_{1s}+ i{\Gamma}}\right|^2,
\label{eq.RXSorb}
\end{equation}
with 
\begin{equation}
 C_{x^2-y^2}=E^{\mbox{\scriptsize out}}_x E^{\mbox{\scriptsize in}}_x
            -E^{\mbox{\scriptsize out}}_y E^{\mbox{\scriptsize in}}_y.
\label{eq.azim}
\end{equation}
Just like Eq.~(\ref{eq.RXSmag}), the polarization dependent part $C_{x^2-y^2}$
is factored out, so that the photon energy dependence becomes independent 
of polarization.

%--------+---------+---------+---------+---------+---------+---------+---------+
\begin{figure}
\includegraphics[width=7cm,keepaspectratio]{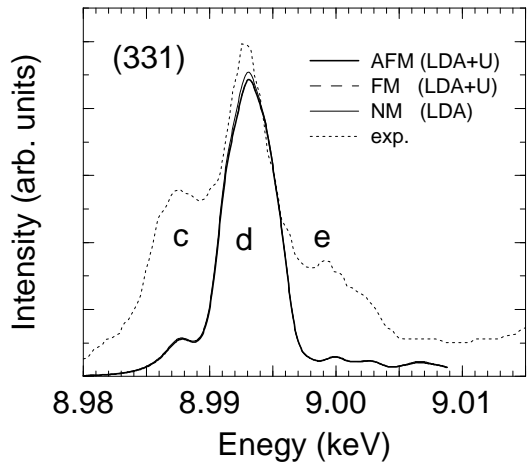}
\caption{Calculated RXS spectra on an orbital superlattice spot $(331)$
in comparison with the experiment.
The thick solid and broken lines show the spectra on the AFM and FM phases
in the LDA$+U$ scheme, respectively. The thin solid line shows the spectra
on the NM phase in the LDA scheme. The dotted line shows the experimental 
data (Ref.[\onlinecite{Caciuffo}]).
The core-hole energy is set the same as in the case of absorption coefficient.
}
\label{fig:RXSorb}
\end{figure}
%--------+---------+---------+---------+---------+---------+---------+---------+
Figure \ref{fig:RXSorb} shows the calculated spectrum for $\vecG=(331)$,
 as a function of photon energy, in comparison with the experiment.  The
 calculation is carried out on both the AFM and FM phases in the LDA$+U$
 scheme, and on the NM phase in the simple LDA scheme.  All three cases of
 calculation give nearly the same result, indicating that the $4p$ states
 are little influenced by the details of the $3d$ states, in agreement of
 the previous calculation of the same kind.  \cite{Caciuffo} In particular,
 the fact that the last case gives the similar result in spite of
 considerably smaller orbital polarizations gives a strong support to the
 mechanism that the $3d$ orbital polarization has little relation to the RXS
 intensity.  This is consistent with previous studies on other
 transition-metal compounds, \cite{Elfimov,Benfatto,Taka1} where the RXS
 intensity are mainly controlled by lattice distortion.  The effect of the
 SOI is negligibly small for the these cases.  The present calculation
 reproduce well the spectra consisting of three peaks, although peaks c and
 e are much smaller than the experimental ones.

The azimuthal angle dependence for $\vecG=(331)$ is explicitly evaluated
from Eq.(\ref{eq.azim}). It is given by
\begin{eqnarray}
 I(\vecG,\omega) &\propto& |\cos\beta\sin 2\psi|^2 \quad 
 {\rm for}\ \sigma\to\sigma'    \nonumber \\
                   &\propto& |\sin\theta\cos\beta\cos 2\psi 
                              + \cos\theta\sin\beta\sin\psi|^2 \nonumber \\
                   &&\quad {\rm for}\ \sigma\to\pi', 
\end{eqnarray}
with $\sin\theta=0.364$, $\tan\beta=3\sqrt{2}c/a=4.02$.
The azimuthal angle $\psi=0$ is again defined such that the $(\bar110)$ axis
is contained in the scattering plane.
The intensity has a period of $2\pi$ not $\pi$ with respect to $\psi$.
This rather complex behavior seems reasonable, since the scattering vector
$\vecG$ is not in a direction of high symmetry.
This is obviously different from the form previously assumed in the analysis
of the experimental data.\cite{Caciuffo}
Figure \ref{fig:azim} shows the peak intensity 
as a function of $\psi$, in comparison with the experiment.
The present form is consistent with the experiment.

\begin{figure}
\includegraphics[width=7cm,keepaspectratio]{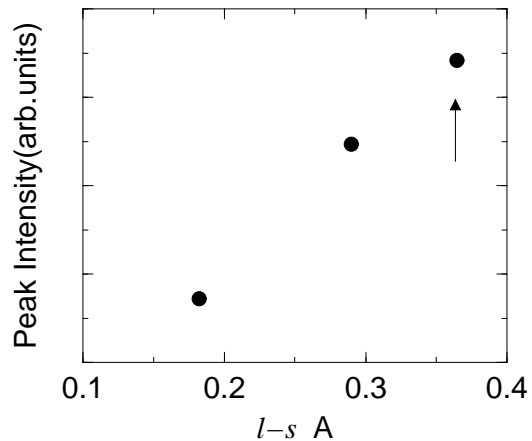}
\caption{Peak intensity of RXS spectra on $(331)$ with varying JTD strength.
The values of $m$ and $\ell+s$ are fixed at $1.962$\AA\ and $4.141$\AA, 
respectively. The circle indicated by an arrow is the result for the actual 
value of $\ell-s$ at room temperature.
}
\label{fig:RXSjtd}
\end{figure}

So far, all the spectra are calculated with the lattice parameters at room
 temperature mentioned before.  If the RXS intensity arises from lattice
 distortion, it increases with increasing JTD strength.  To confirm this, we
 calculate the intensity with changing the JTD strength.  Figure
 \ref{fig:RXSjtd} shows the intensity of the peak at $\hbar\omega=8995$ eV
 as a function of $\ell-s$ (the values of $m$ and $\ell+s$ are kept to be
 $1.962$\AA\ and $4.141$\AA, respectively).  The intensity monotonically
 increases with increasing values of $\ell-s$, consistent with the above
 observation. It has been observed in the experiment that the intensity
 increases by factor 2 with temperature going through the magnetic phase
 transition temperature, and this behavior has been interpreted as a
 consequence of a strong coupling between orbital and spin degrees of
 freedom.  \cite{Paolasini} The result that the spectra are mainly
 controlled by lattice distortion indicates that the lattice distortion
 becomes large in the magnetic phase.  The direct determination of lattice
 parameters have not been carried out yet. 
%--------+---------+---------+---------+---------+---------+---------+---------+
%--------+---------+---------+---------+---------+---------+---------+---------+

\section{Concluding Remarks}

We have calculated the RXS spectra around the $K$ edge of Cu in KCuF$_3$ on
 the basis of the {\em ab initio} calculation, that is, the FLAPW method in the
 LDA$+U$ scheme.  The lattice distortion is introduced as inputs of
 calculation.  We have obtained finite intensities on magnetic superlattice
 spots by taking account of the SOI.  The intensities are about three orders
 of magnitude smaller than those on orbital superlattice spots, consistent
 with the experiment.  This is the first result to have evaluated the
 magnetic RXS spectra in the {\em ab initio} level.  Since the intensity
 disappears without the SOI, the present result indicates that the spectra
 arises from the orbital polarization, not from the spin polarization, in
 the $4p$ states.  On the other hand, the spectra change little with turning
 off the SOI from the $3d$ states.  This indicates that the $3d$ orbital
 moment has no role to induce the $4p$ orbital polarization.  This situation
 is different from our previous finding in the study of the magnetic RXS in
 CoO and NiO that the $4p$ orbital polarization is induced by the $3d$
 orbital moment through the mixing of the $4p$ states with the $3d$ states
 of {\em neighboring} Cu atoms and the intra-atomic $4p$-$3d$ Coulomb
 interaction.\cite{Iga1,Iga2} This difference may come from the fact that
 the $3d$ orbital moment is about 1/3 of that in NiO and 1/10 of that in
 CoO.  Furthermore, the mixing effect may become smaller than in NiO and CoO,
 since neighboring Cu sites are always intervened by F atoms.  Closely
 related is the phenomenon of the magnetic circular dichroism (MCD) in the
 $K$-edge absorption in ferromagnetic metals Fe, Co, Ni.\cite{Schutz} In this
 case, the $4p$ orbital polarization was found to be induced by the $3d$
 orbital moment through the $4p$ mixing with the $3d$ states at {\em
 neighboring} sites.\cite{Iga3,Iga4}

We have also calculated the RXS intensities on orbital superlattice spots.
  The spectra are independent of whether the system is in the AFM phase or
 in the FM phase, and of the orbital polarization in the $3d$ states.  We
 have explicitly shown that the spectra are mainly controlled by the lattice
 distortion through the calculation with varying JTD strength.  The present
 result is consistent with the previous studies of the RXS on
 transition-metal oxides,\cite{Elfimov,Benfatto,Taka1,Taka2,Taka3} but shows
 contrast with the RXS around the $L_{\rm III}$ edge in the quadrupole
 ordering phase of the rare-earth compound CeB$_6$, where the spectra seems
 directly controlled by the quadrupole order.  \cite{Nakao01,Nagao,Nagao2}

\begin{acknowledgments}
We have used the FLAPW code developed by N. Hamada.  We
 thank him for allowing to use his code.  This work was partially supported
 by a Grant-in-Aid for Scientific Research from the Ministry of Education,
 Culture, Sports, Science, and Technology, Japan.
\end{acknowledgments}

\bibliography{p3}

\newpage
%
%--------%---------%---------%---------%---------%---------%---------%---------%--------%---------%---------%---------%---
%
\end{document}